\documentstyle[cjaa,epsfig,11pt]{article}

\newcommand{\bq}{\begin{equation}}
\newcommand{\eq}{\end{equation}}

\lefthead{H.G. Wang et al.} \righthead{Energy loss of particles
in Pulsar Radio Emission Region}
\begin{document}

\title{Energy Loss of Particles in Pulsar Radio Emission Region}
\author{H.G. Wang, Y. Wang, G.J. Qiao, R.X. Xu\\
        CAS-PKU joint Beijing Astrophysical Center
        and Department of Astronomy, \\
        Peking University, Beijing 100871, China}

\begin{abstract}

Extremely relativistic particles play an important role in various
radiative mechanisms of pulsar radio as well as high-energy
emission. It is thus very essential to estimate the Lorentz factor
of these particles in the radiation regions both observationally
and theoretically. In this paper, according to the frequency
dependence of component separation and considering some radiation
processes, we obtain the Lorentz factor and its variation in the
radiation region observationally for seven multi-frequency
observed pulsars. It is found that the Lorentz factor decays
substantially in the emission region, and the decay can not be
well understood by all known energy loss processes. This result
hints that there are some unknown processes which are responsible
to loosing energy effectively.

\end{abstract}


\section{Introduction}\label{Sect1}

The energy loss of relativistic particles in pulsar magnetosphere
have been studied by many authors (Daugherty \& Harding 1989,
Sturner 1995, Zhang et al. 1997, Lyubarskii \& Petrova 2000,
hereafter LP2000). Three possible energy loss mechanisms, i.e.,
inverse Compton scattering (ICS) of thermal photons by
relativistic particles, triplet pair production and curvature
radiation of particles, were investigated by Sturner (1995). It is
found that the resonant ICS process in strong magnetic fields is
the dominant energy loss mechanism when the Lorentz factors are
between 10 and $10^4$ and pulsar surface temperature is $\sim
10^6$ K. Considering resonant ICS of thermal photons by secondary
particles ($10<\gamma<10^4$) above the pulsar inner gap, LP2000
found that the ratio of Lorentz factor to its initial value,
$\gamma/\gamma_0$, decreases generally to $40\%$ within a height
$\lesssim10$ km along the magnetic field lines above the polar cap
and nearly keeps constant at higher distances.

The Lorentz factor $\gamma$ can be estimated from some
observational properties (e.g. microstructure width) of pulsars.
Lange et al. (1998) assumed that the widths of microstructure are
related to the beam widths of relativistic particles. They found
that the derived $\gamma$, according to the observed micropulse
typical widths of PSR B0950+08, PSR B1133+16 and PSR B2016+28, is
in a range from 240 to 520 if the angle between the line of sight
and pulsar rotation axis is $90^{\circ}$. $\gamma$ could be even
larger since this angle may be less than $90^{\circ}$ (Lyne \&
Manchester 1988, hereafter LM88, Rankin 1993).

However, as far as we know, there are few works trying to derive
energy loss of particles  in the emission region observationally.
Thorsett (1991, hereafter T91) studied the dependence of component
separation on observing frequency for seven pulsars. It is found
that the separation of emission components varies with observing
frequencies smoothly. According to this observational analysis and
employing  pulsar radio emission models, we calculate the $\gamma$
factors at different heights. It is found that in the radio
emission region, the Lorentz factor decays significantly, which
can not be understood by the energy loss processes known hitherto.

In $\S 2$ the observational facts of the frequency dependence of
component separation  are introduced. The basic assumption,
calculation and results are showed in $\S3$. In $\S 4$,
conclusions and discussions are presented.

\section{The frequency dependence of component separation}

It is based on the dependence of component separation
$\Delta\theta$ on observing frequencies $\nu$ to find out the
energy distribution of particles in the emission region of
pulsars. The dependence has been studied for a long time. Earlier
analysis found that a simple power law could not fit the frequency
dependence of component separation (hereafter FDCS) for most
pulsars (Lyne et al. 1971, Sieber et al.1975, Rankin 1983, Slee et
al. 1987). Traditionally two power laws were introduced,
%
\bq \Delta\theta=\left\{\begin{array}{cc} A_1\nu^{-\alpha_1},&
\left(\nu<\nu_{\rm b}\right),\\
A_2\nu^{-\alpha_2}, & \left(\nu>\nu_{\rm b}\right),
\end{array}
\right. \eq
where the break frequency $\nu_{\rm b}$ is usually between
0.14--1.5GHz, $\alpha_2<\alpha_1$ indicating that at higher
frequencies the component separation is less frequency dependent.
Rankin (1983) even introduced a third power-law at intermediate
frequency to fit PSR B1133+16. Collecting all available data, T91
restudied the FDCS for seven pulsars that exhibit double- or
multiple-component average profiles, and proposed a smooth
function,
%
\bq \Delta\theta=A\nu^{-n} +\Delta\theta_{\rm min}, \eq
to model for each pulsar (for multiple-component profiles, the
component separation is the width between the two peaks of outmost
cone). As T91 pointed out, it was mainly the insufficient sampling
that led those earlier works to report a break frequency. Similar
to the FDCS, the average profile width measured at 50\% or 10\% of
component peaks also narrows as frequency increases (Rankin 1983,
Xilouris et al. 1996).

Our following analysis and results are based on Eq.(2). On the one
hand, the index $n$ and minimum component separation
$\Delta\theta_{\rm min}$ were given by T91. On the other hand,
$\Delta\theta$ is a geometrical result which depends on emission
height, and $\nu$ is model-dependently related to the Lorentz
factor in the emission region. Combining these two hands, we can
derive a relationship between the distances and the Lorentz
factors, when the geometric effect and a radio emission model are
involved.

\section{Results: Electron Lorentz factor and emission distance}


Two assumptions are used in our analysis: 1) the magnetic field
lines are dominantly dipolar in pulsar radio emission region, 2)
radio emission arises from the last open field lines.
The beaming angle $\theta_\mu$ (the angle between the tangent of
magnetic field and the magnetic axis) can then be calculated from
the viewing geometry (LM88),
%
\bq \sin^{2}\left(\theta_\mu\over2\right)=
\sin^2\left(\Delta\theta\over 4\right)\sin\alpha
\sin\left(\alpha+\zeta\right)+\sin^2\left(\zeta\over2\right), \eq
where the impact angle $\zeta$ is the angle between the line of
sight and magnetic axis, $\alpha$ is the inclination angle.
$\alpha$ and $\zeta$ can be obtained from observations (LM88,
Rankin 1993). Geometrically, one has the following equations (Qiao
\& Lin 1998, hereafter QL98) about $\theta_\mu$, from which the
distance $r$ can be calculated,
%
\bq
\tan\theta_\mu={{3\sin\theta\cos\theta}\over{(2-\sin^2\theta)}},
\eq
%
\bq \sin^2\theta={r\over R_{\rm LC}}, \eq
where the polar angle $\theta$ is the angle between position
vector ${\bf r}$ (from neutron star center to emission point) and
the magnetic axis, $r=|{\bf r}|$, $R_{\rm LC}={{cP}\over{2\pi}}$
is the radius of light cylinder, with $c$ the light velocity and
$P$ the pulsar rotating period.

The Lorentz factor $\gamma$ of relativistic particles can be
model-dependently obtained for a given $r$. Among pulsar radio
emission models, the curvature radiation model (e.g., Ruderman \&
Sutherland 1975, hereafter RS75) and the ICS model (Qiao 1988,
QL98) predict different functions of $\nu=\nu(\gamma,r)$, which
will be used below to calculate $\gamma=\gamma(r)$, respectively.

In the RS75 model, relativistic particles give out emission
through curvature radiation with frequency
%
\bq \nu={3\over2} \gamma_{\rm CR}^3
\left({c\over{2\pi\rho}}\right), \eq
then,
%
\bq \gamma_{\rm CR}=\left({4\pi\rho \nu}\over {3c}\right)^{1/3},
\eq
where `CR' denotes `curvature radiation', $\rho$ is the curvature
radius at a given point, which reads
%
\bq
\rho={{(1+3\cos^2\theta)^{3/2}\sin\theta}\over{3(1+\cos^2\theta)}R_{\rm
LC}}. \eq

In the ICS model, low frequency electromagnetic wave, with
frequency of $\nu_0\sim 10^6$Hz, is produced by inner gap sparking
(though the frequency may vary for different pulsars due to
different gap parameters, it does not change the results
essentially for a given pulsar, so we simply take the frequency as
$\nu_0\sim 10^6$Hz). These low-frequency photons are up-scattered
by relativistic electrons to higher frequencies (QL98),
%
\bq \nu={3\over2} \gamma_{\rm ICS}^2 \nu_0
\left(1-\beta\cos\theta_{\rm i}\right), \eq
then one has
%
\bq \gamma_{\rm ICS}=\left(2\nu\over
{3\nu_0}\right)^{1/2}\left(1\over {1-\beta \cos\theta_{\rm
i}}\right)^{1/2} \eq
where $\beta=(1-1/\gamma_{\rm ICS}^2)^{1/2}$, the incident angle
$\theta_{\rm i}$ is the angle between the wave vector of low
frequency wave and the direction of electron moving along field
lines, which can be calculated to be (QL98),
%
\begin{equation}
\cos\theta_{\rm i}={2\cos\theta+(R/r)(1-3\cos^2\theta)
\over \sqrt{(1+3\cos^2\theta)[1-2(R/r)\cos\theta+(R/r)^2]}}.%
\label{thetai}
\end{equation}

Employing the observational parameters listed in column 2 to 7 in
Table 1, viz., the frequency range, $A$, $n$, and
$\Delta\theta_{\rm min}$ given by T91, $\alpha$ and $\zeta$ given
by LM88, we  calculate $\gamma$ and $r$ numerically and present
the data in Fig. 1 for each pulsar. The values of minimum distance
$r_{\rm min}$, maximum Lorentz factors $\gamma_{\rm CR,max}$ and
$\gamma_{\rm CR,max}$, and the ratio of $r_{\rm max}\over r_{\rm
min}$, $\gamma_{\rm CR,min}\over \gamma_{\rm CR,max}$ and
$\gamma_{\rm ICS,min}\over \gamma_{\rm ICS,max }$ are listed in
table 1, where $r_{\rm min}$ and $\gamma_{\rm
CR,max}$($\gamma_{\rm ICS,max}$) are obtained from the maximum
observable frequency, while $r_{\rm max}$ and $\gamma_{\rm
CR,min}$($\gamma_{\rm ICS,min}$) are from the minimum observable
frequency.

\section{Conclusions and discussions}

According to the frequency dependence of component separation of
seven pulsars (T91), we obtain a model-dependent relationship
between the Lorentz factors $\gamma$ of extremely relativistic
particles and the emission distances $r$ for each pulsar. From the
analysis, we have following two points.

1). The radio emission of outmost cone typically arise from the
distance of about $100\rm km$ to about $300\rm km$, which is
consistent with the previous results for these pulsars (e.g.
Phillips 1992, Hoensbroech \& Xilouris 1997).

2). Either $\gamma_{\rm CR}$ or $\gamma_{\rm ICS}$ decreases
monotonically as $r$ increases. Within the range of $r$,
$\gamma_{\rm CR}$ may reduce by about 60\% and 90 \% with respect
to their maximum values, respectively. The results reflect that
particles undergo severe energy loss in the radio emission region,
which is essentially inconsistent with the theoretical prediction
made by many authors, for example, LP2000, as will be discussed
below.

Some discussions related to our analysis are as follows.

1). In our analysis, emission is produced at the last open field
lines. In fact, to derive $\gamma (r)$, one can also assume that
radio emission arises from inner magnetic fields, and the results
are nearly the same, for an example, with the radius $R_{\rm
e}=\lambda R_{\rm LC}$, where the factor $\lambda>1$ (this means
that the field lines are closer to to the magnetic axis, see
QL98). Fig. 2 compares the curves of $\gamma (r)$ obtained for the
last open field lines and inner field lines with $\lambda=1.2$
(corresponds to the field lines satisfying $\theta=0.91\theta_{\rm
P}$, where $\theta_{\rm P}$ is the polar cap angle defined by the
last open filed line at pulsar surface) for PSR B 0525+21, from
which one can see that the $\gamma (r)$ does not show significant
difference. The uncertainty of inclination angle $\alpha$ and
impact angle $\zeta$ may influence the $\gamma (r) $, as shown in
Fig. 3, which presents the $\gamma (r) $ curves for three groups
of $\alpha$ and $\zeta$ , i.e., ($23.2^\circ$, $0.7^\circ$) (given
by LM88), ($23.2^ \circ$, $2.0^\circ$) and ($90^\circ$,
$0.7^\circ$).
 Here It shows that the uncertainty of $\alpha$ may change the results
of $r$ greatly, but in general, $\gamma (r)$ does not change
essentially.

2). Interestingly, our results show that there must be some
additional energy-loss processes in the radio emission regions.
But so far we did not find any reasonable effective mechanism for
the energy loss. Theoretical study, e.g., LP2000, shows that
$\gamma$ nearly does not decay at the emission region higher than
10km if only electron ICS process of the thermal photons from
neutron star surface is included. Also pulsar radio emission can
not lead electrons loose  energy significantly (Lin 1997).

A possible effective process may be the resonant ICS of soft
X-rays produced in outer gap with the relativistic particles.
Cheng \& Zhang (1999) present a model of multi-component X-ray
emission from the rotation-powered pulsars, in which hard thermal
and soft thermal X-rays originate from the back-flow current of
the outer gap. The directions of the soft X-rays are diffused, so
that the resonant ICS may be effective due to  proper incident
angles. However, there is a criterion to determine whether a
pulsar has an outer gap or not (Zhang \& Cheng 1997, Cheng \&
Zhang 1999). The size of the outer gap, which is the ratio between
the potential drop of the outer gap and the total potential of the
open field lines, limited by the soft thermal X-rays from the
pulsar surface, can be determined as,
%
\bq f=5.5P^{26/21}B_{12}^{-4/7}, \eq
where $P$ is the rotation period of pulsars in seconds, $B_{12}$
the surface magnetic field strength in $10^{12}$ G. Only when
$f\leq 1$ can the outer gap exist. Unluckily, we find $f>1$ for
all the seven pulsars, as presented in Table 2, and the outer gap
may not exist.
It seems that the criterion of Zhang \& Cheng needs to be
re-investigated, or other energy loss processes should be
proposed.

3). It should be emphasized that our result is model dependent.
According to other radio emission models, e.g., relativistic
plasma emission models (Melrose \& Gedalin 1999 and references
therein), some other interesting constrains on the relativistic
particles or plasma may be obtained. But in this paper we just
concentrate on the particle Lorentz factor so that only the
Curvature radiation model and ICS model are employed.


\acknowledgements We are grateful to  Dr. B. Zhang, Dr. J.L. Han
and other members of the pulsar group for helpful discussion. This
work is partly supported by NSF of China, the Climbing project,
the National Key Basic Research Science Foundation of China, and
the Research Fund for the Doctoral Program Higher Education.

\newpage\eject

\begin{table}
\leftskip -0.8in
\caption{Parameters given by T91 and our results}
\begin{tabular}{lllllllllllll}\hline

\textbf{\small PSR B} &\textbf{\small Freq. Range} &\textbf{$A$}
&\textbf{$\Delta\theta_{\rm min}$} &\textbf{$n$}
&\textbf{$\alpha$} &\textbf{$\zeta$} &\textbf{\small $r_{\rm
min}$}&\textbf{\small $\gamma_{\rm CR,max}$} &\textbf{\small
$\gamma_{\rm ICS,max}$} &\textbf{$r_{\rm max}\over r_{\rm min}$}
&\textbf{$\gamma_{\rm CR,min}\over \gamma_{\rm CR,max}$}
&\textbf{$\gamma_{\rm ICS,min}\over \gamma_{\rm ICS,max
}$} \\

&\textbf{\small (GHz)} & &\textbf{\small ($^\circ$)} &\
&\textbf{\small ($^\circ$)}
&\textbf{\small ($^\circ$)} &\textbf{\small (km)} \\
\hline
0301+19 &0.1-2.7 &86 &0.9 &0.34 &31.9 &1.8 &59 &463 &3763 &4.4 &0.43 &0.092\\
0329+54 &0.1-10.7 &1059 &19.8 &0.96 &30.8 &2.9  &169 &781 &2957 &2.2 &0.24 &0.068\\
0525+21 &0.05-4.9 &90 &9.5 &0.47 &23.2 &0.7 &132 &762 &6105 &4.2 &0.28 &0.048\\
1133+16 &0.026-10.7 &53 &4.4 &0.50 &51.3 &3.7  &135 &819 &4062&2.8 &0.16 &0.031\\
1237+25 &0.08-4.9 &79 &7.9 &0.52 &48.2 &0.9  &106 &621 &3419&3.1 &0.31 &0.076\\
2020+28 &0.1-14.8 &1103 &9.1 &1.08 &71.2 &3.6  &71 &666 &2374&2.3 &0.22 &0.063\\
2045-16 &0.08-4.9 &45 &7.9 &0.36 &36.7 &1.1  &132 &684 &4007 &2.7 &0.30 &0.078\\
\hline
\end{tabular}

\end{table}

\begin{table}
\caption{numerical values of $f$}
\begin{tabular}{llll}\hline

\textbf{PSR B} & \textbf{$P$ (s)} & \textbf{$B_{12}$ (G)}
& \textbf{$f$}\\
\hline
0301+19&1.388&1.34&6.98\\
0329+54&0.715&1.21&3.25\\
0525+21&3.746&12.25&6.74\\
1133+16&1.188&2.11&4.44\\
1237+25&1.382&1.15&7.58\\
2020+28&0.343&0.81&1.66\\
2045--16&1.962&4.64&5.27\\
\hline
\end{tabular}
\end{table}

\begin{figure}
\setlength{\abovecaptionskip}{0pt}
\setlength{\belowcaptionskip}{0pt}%
\centering
\includegraphics[width=15cm,height=18cm]{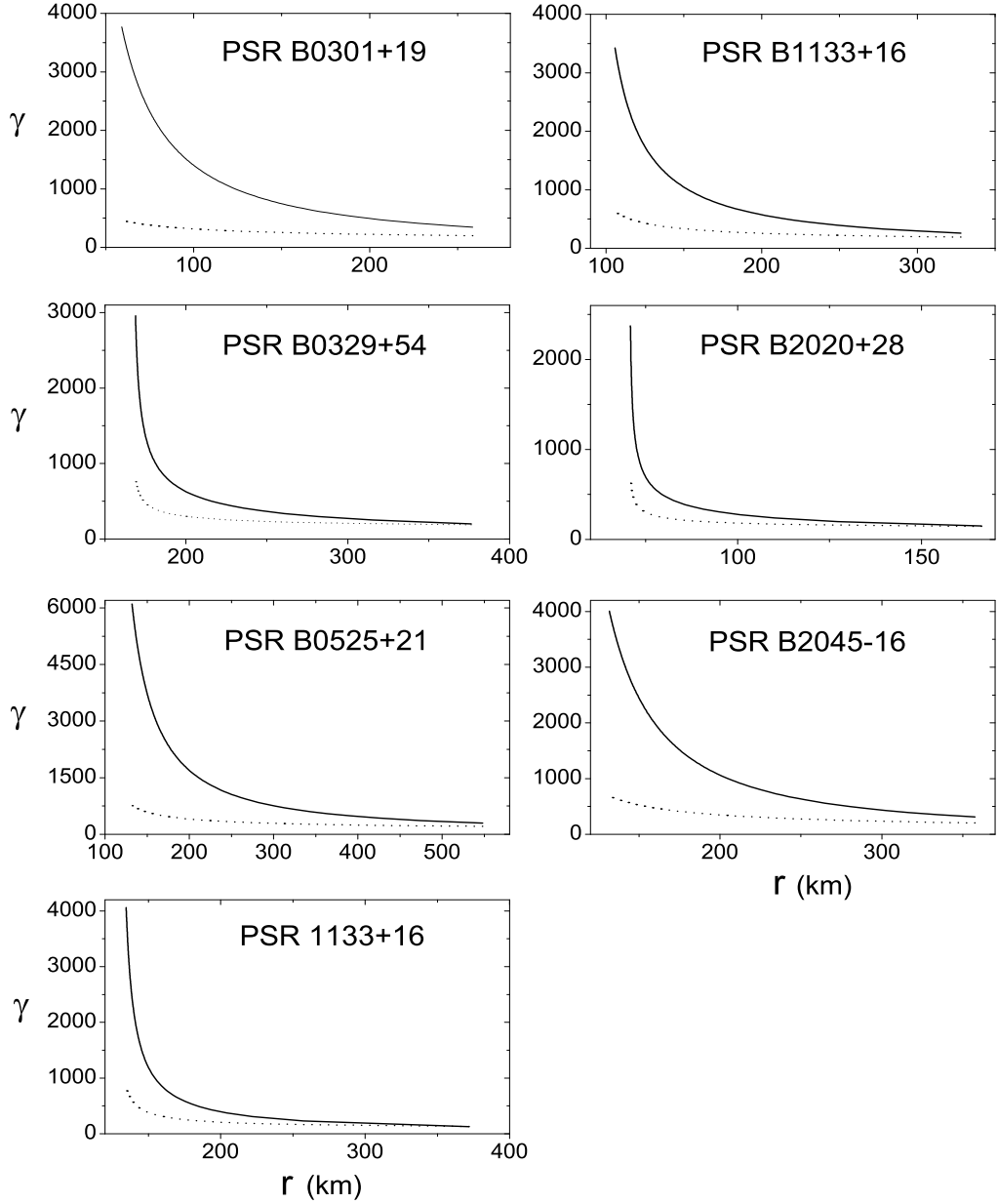}
\caption{
The $\gamma(r)$ relationships for seven pulsars. The dot and solid
curves represent the data calculated for curvature radiation and
inverse Compton scattering respectively.
}%
\label{figl}
\end{figure}
\begin{figure}
\setlength{\abovecaptionskip}{0pt}
\setlength{\belowcaptionskip}{0pt}%
\centering
\includegraphics[width=8cm,height=7cm]{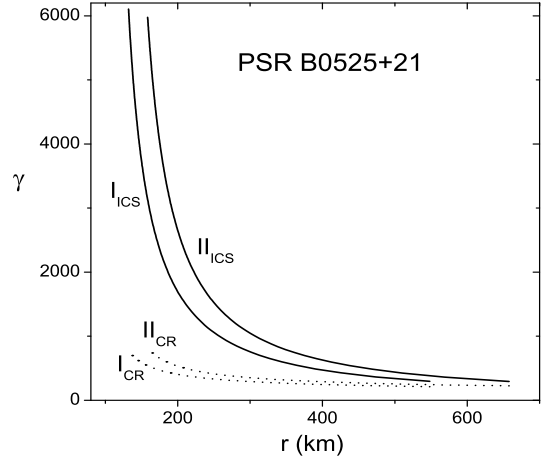}
\caption{
The effect of inner field lines to $\gamma(r)$. The curves marked
by I are calculated for the last open field lines, and II for the
inner field lines with $\lambda=1.2$. `CR' denotes curvature
radiation and `ICS' denotes inverse Compton scattering.
}%
\label{fig2}
\end{figure}
\begin{figure}
\setlength{\abovecaptionskip}{0pt}
\setlength{\belowcaptionskip}{0pt}%
\centering
\includegraphics[width=8cm,height=7cm]{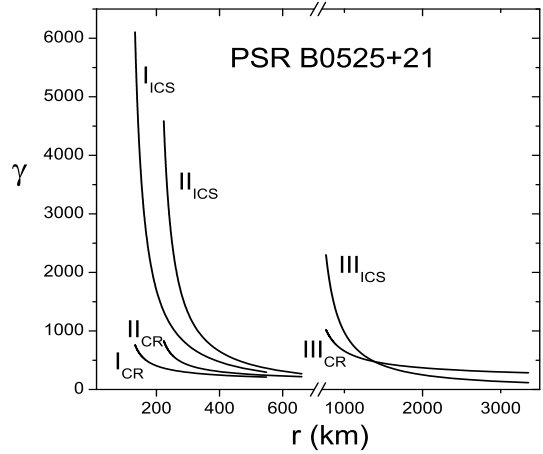}
\caption{
The influence of inclination angle $\alpha$ and impact angle
$\zeta$ to $\gamma(r)$. The curves marked by I are calculated for
($23.2^\circ$, $0.7^\circ$), II for ($23.2^ \circ$, $2.0^\circ$)
and III ($90^\circ$, $0.7^\circ$). The marks of `CR' and `ICS' are
the same as those in Fig.2.
}%
\label{fig3}
\end{figure}

\end{document}